\documentstyle[12pt,preprint]{aastex}

\shorttitle{Image Masks}
\shortauthors{Kuchner}
\begin{document}

\slugcomment{submitted to ApJ January 13, 2004}

\title{A Unified View of Coronagraph Image Masks}
\author{Marc J. Kuchner\altaffilmark{1}}
\affil{Princeton University Observatory \\ Peyton Hall, Princeton, NJ 08544}
\altaffiltext{1}{Hubble Fellow}
\email{mkuchner@astro.princeton.edu}

\begin{abstract}

The last few years have seen a variety of new image mask designs
for diffraction-limited coronagraphy.  Could there still be useful designs
as yet undiscovered? To begin to answer this question, I survey and
unify the Fraunhofer theory of coronagraph image masks in the context
of a one-dimensional classical coronagraph.  I display a complete
solution to the problem of removing on-axis light assuming
an unapodized entrance aperture and I introduce the attenuation function, a measure
of a generic coronagraph's off-axis performance.  With these tools, I demonstrate
that the masks proposed so far form a nearly complete library of image
masks that are useful for detecting faint extrasolar planets.

\end{abstract}


\section{INTRODUCTION}

In a classical coronagraph \citep{lyot39}, an image mask placed at a
telescope's focus and a Lyot stop in a succeeding pupil plane block most of the light
from a bright on-axis source allowing the telescope to better image faint off-axis sources.
Classical coronagraphs have provided the first images of a brown dwarf orbiting a nearby star
\citep{naka95} and striking images of circumstellar disks \citep[e.g.][]{smit84, clam03} and
promise to enable imaging of bright extrasolar planets within the next decade \citep{trauger03}.
Initial studies suggest that classical coronagraphs can potentially image even extrasolar
terrestrial planets \citep[see the review by][]{ks03}.

The last few years have seen a cornucopia of new coronagraph designs, including a variety
of new image mask designs, typically invented using Fraunhofer diffraction theory. Perhaps
construction tolerances and vector electromagnetic effects will limit the practicality of all these designs.
However, the more choices of masks we have explored using Fraunhofer theory, the more likely it
seems we may find a useful one for terrestrial planet finding.  With extrasolar terrestrial planets
in mind, I attempt to develop a unified picture of coronagraph image masks to determine the range
of potentially useful solutions to the Fraunhofer diffracted-light problem.  I concentrate first on
a one-dimensional coronagraph with a tophat entrance aperture
and then consider more general entrance apertures.

\section{A SIMPLE CORONAGRAPH}

We will examine a one-dimensional coronagraph \citep{siva01} comprising
an entrance aperture, $A$, an image mask, $\hat M$, and a Lyot stop, $L$,
each of which is represented by a complex-valued function constrained to have
absolute value $\le 1$.
We use the notational conventions of \citet{kuch02} and \citet{kuch03};
letters with hats represent pupil plane quantities, and quantities transform as follows:
\begin{equation}
M(u)=\mbox{FT}[\hat M(x)]=\int_{-\infty}^{\infty} \hat M(x) \, e^{-2 \pi iux} \, dx.
\end{equation}
We will assume the optics have removed any quadratic phase terms associated with small
focal lengths.  We will set all time-varying factors in the electric fields equal to 1; whenever we say
``field'', we mean the amplitude of the time varying field.
Mostly we will discuss monochromatic effects, though sometimes we will consider
the effects of finite bandwidth.

We will use dimensionless units, but our units can be translated into
physical distances as follows.  In the pupil planes, our coordinates ($u$, $u_1$) measure
distance from the optical axis in units of the
local pupil diameter, which for example, at the primary mirror would simply
be $D$, the diameter of the primary mirror.  However, since the model is one dimensional,
the primary mirror doesn't have a diameter so much as a width; it extends forever in one direction.
In the image planes, our coordinates ($x$, $\theta$, etc.) measure distance
from the optical axis in units of the diffraction scale, which is ordinarily
$\lambda f$, where $\lambda$ is the wavelength of light and $f$ is the focal ratio.
In the plane of the sky, they measure angle from the optical axis in units of $\lambda/D$ radians.

An incoming wave incident on the entrance aperture creates a field with amplitude $E(u)$.
In our model, when an incoming wave interacts with a stop or mask, the function
representing the mask multiplies the wave's complex amplitude.  So after the wave interacts with
the entrance aperture, the amplitude becomes $A(u)\cdot E(u)$.

After interacting with the entrance aperture, the beam propagates to an image plane, where the new field amplitude is the Fourier transform
of the pupil plane field amplitude, $\hat A(x) * \hat E(x)$, where $*$ denotes convolution.
There the beam interacts with the image mask, and the field amplitude becomes $\hat M(x) \cdot (\hat A(x) * \hat E(x))$.
Then the beam propagates back to a second pupil plane, where the field amplitude is $M(u) * (A(u) \cdot  E(u))$.

In the second pupil plane, the wave interacts with a Lyot stop, changing
the field amplitude to $L(u) \cdot [M(u) * (A(u) \cdot E(u))]$.
Then the beam propagates to a final image plane, where the final image
is $\hat L(x) * [\hat M(x) \cdot (\hat A(x) * \hat E(x))]$.
For a point source, the intensity of the final image is proportional to the
absolute value of this quantity squared.


A coronagraph aims to reduce the final image of an on-axis point source, for which $E(u)$ is a constant,
which we can set equal to 1.
The field created by such a source in the second pupil plane, after the Lyot stop,
is $L(u) \cdot (M * A)$ \citep{siva01}.
The corresponding final image field
is the Fourier transform of this quantity, $\mbox{FT}(L \cdot (M * A))$.
Masks for which
\begin{equation}
[L \cdot (M * A)](u) = 0
\label{eq:zero}
\end{equation}
will block identically all of the light from an on-axis point source; we aim to find these masks.
This paper concerns solutions and approximate solutions to this linear problem.

\subsection{Tophat Entrance Aperture, Arbitrary Lyot Stop}

We will examine first a system where the entrance aperture, $A(u)$, is opaque ($A(u)=0$)
for $|u| > 1$ and transparent ($A(u)=1$) for $|u|< 1/2$, and the Lyot stop, $L(u)$, is
opaque for $|u| > (1-\epsilon)/2$, where $\epsilon \le 1$.
To understand Equation~\ref{eq:zero} in the context of our simple coronagraph,
we can write $A(u)$ as a difference of two Heaviside step functions, $H(u)$:
\begin{equation}
A(u)=H(u+1/2)-H(u-1/2)
\end{equation}
Then, since convolution with a Heaviside step function represents indefinite
integration, we can write
\begin{equation}
M(u)*A(u)={\cal M}(u+1/2)-{\cal M}(u-1/2)
\label{eq:ma}
\end{equation}
where $(d/du){\cal M}(u)=M(u)$.
Now we can see that Equation~\ref{eq:zero} demands only that
\begin{equation}
{\cal M}(u+1/2) = {\cal M}(u-1/2) \quad \mbox{for $-(1-\epsilon)/2 < u < (1-\epsilon)/2$}
\end{equation}
or equivalently.
\begin{equation}
{\cal M}(u) = {\cal M}(u-1) \quad \mbox{for $\epsilon/2 < u < 1-\epsilon/2$}.
\label{eq:demandsonlythat}
\end{equation}

An uncountable infinity of non-trivial masks meet this symmetry criterion.
Figure~\ref{fig:calm}a shows a generic function, ${\cal M}(u)$ that satisfies
Equation~\ref{eq:demandsonlythat}.  In the region $\epsilon/2 < u < 1-\epsilon/2$,
${\cal M}(u)$ looks the same as it does in the region $-\epsilon/2 < u < -(1-\epsilon/2)$;
outside those regions, the function shown in Figure~\ref{fig:calm}a has some
random squiggles, to remind you that
Equation~\ref{eq:demandsonlythat} doesn't specify anything about the function there.

\subsection{Notch Filter Masks and Band-Limited Masks}

\begin{figure}
\epsscale{0.85}
\plotone{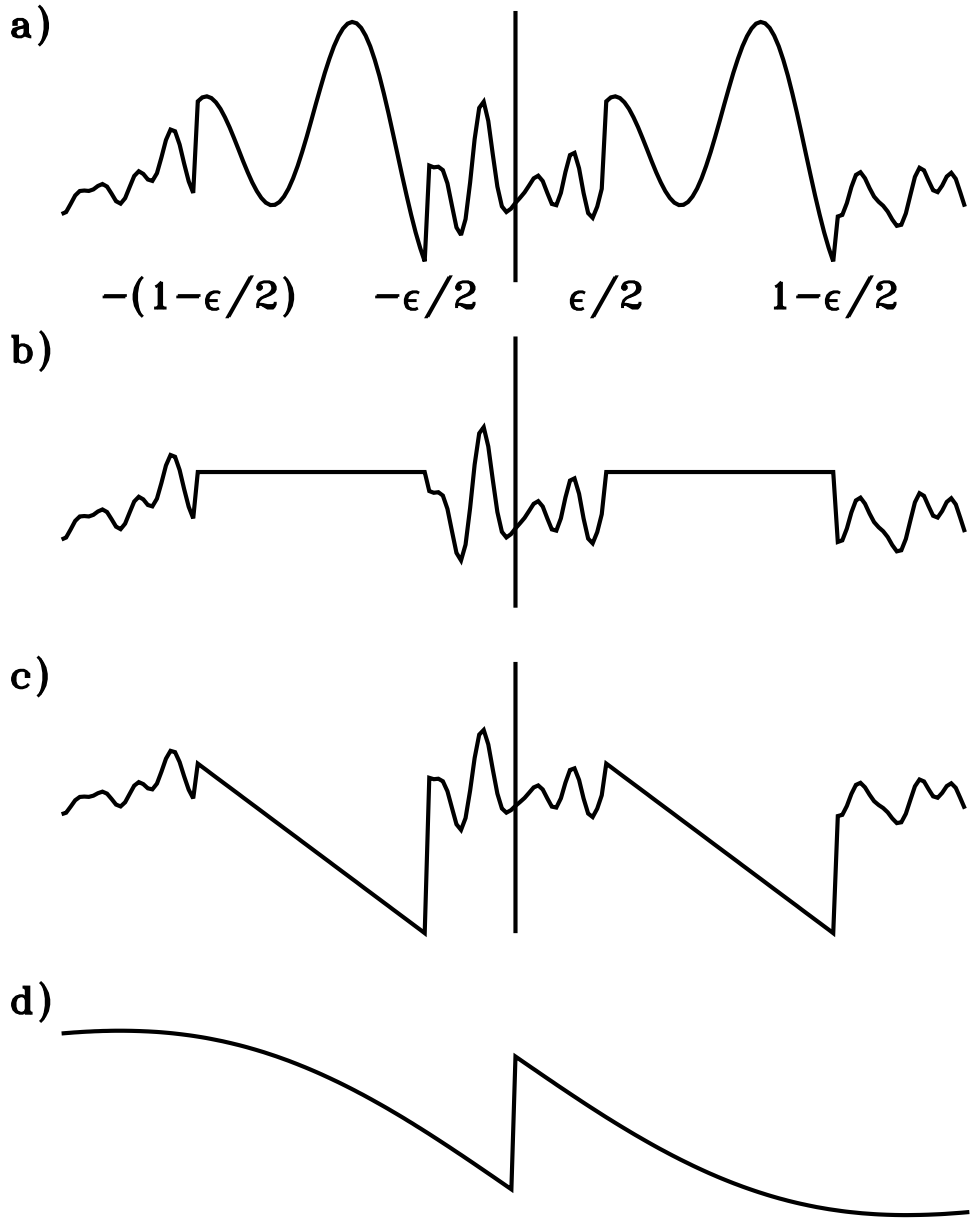}
\caption{${\cal M}(u)$ for a) a generic solution of Equation~\ref{eq:demandsonlythat}, b) a zeroth order, or notch filter mask
c) a first order mask d) a 1-D disk phase mask, a good approximation to a first order mask.   Eliminating on-axis light
in a 1-D coronagraph with tophat entrance aperture requires finding a function with this
translational symmetry. \label{fig:calm}}
\end{figure}

Figure~\ref{fig:calm}b shows the simplest interesting solution to Equation~\ref{eq:demandsonlythat}:
\begin{equation}
{\cal M}(u) \equiv \mbox{constant} \quad \mbox{for $\epsilon/2 < |u| < 1-\epsilon/2$}.
\label{eq:notchcalm}
\end{equation}
This statement about ${\cal M}(u)$ translates into two requirements on
$M(u)$:
\begin{equation}
M(u)=0 \quad \mbox{for $\epsilon/2 < |u| < 1-\epsilon/2$}
\label{eq:notch1}
\end{equation}
and
\begin{equation}
\int_{-\epsilon/2}^{\epsilon/2} M(u) \, du =0.
\label{eq:notch2}
\end{equation}
We can call functions $\hat N(x)$ whose Fourier transforms, $N(u)$, satisfy
Equation~\ref{eq:notch1} ``notch filter'' functions.
Masks satisfying Equation~\ref{eq:notchcalm} or equivalently, both Equations~\ref{eq:notch1}
and \ref{eq:notch2}, are called notch filter masks \citep{kuch03, debe04}.

Notch filter masks for which
$M(u)=0$ for $|u| > \epsilon/2$ are called band-limited masks \citep{kuch02}.
Band-limited masks are an important subset of notch-filter masks
because as we will see below, only the
low-wavenumber components of any mask affect the off-axis light.
Naturally, band-limited functions display wide variation, from the $\hat M(x)=\sin^2(\pi \epsilon x/2)$ mask
that optimizes inner working angle, to masks where $\hat M(x)$ is a prolate spheroidal
wavefunction \citep{kasd03}.  These latter masks optimize search area.  The Gaussian mask \citep{siva01} is
a good approximation to the prolate-spheroidal wavefunction band-limited mask.

\subsection{First Order Masks}

The next simplest solution to Equation~\ref{eq:demandsonlythat}
is one that adds a linear slope to ${\cal M}(u)$, while maintaining the needed translational
symmetry. I.e.,
\begin{equation}
{\cal M}(u) \equiv \left\{  \begin{array}{ll}
            k_1 u &\mbox{for $\epsilon/2 < u < 1-\epsilon/2$} \\
            k_1 (u+1) &\mbox{for $-(1-\epsilon/2) < u < -\epsilon/2$}. \end{array}
	    \right.
\end{equation}
Figure~\ref{fig:calm}c shows an example of such a sawtooth-like function.
When ${\cal M}(u)$ takes this form,
\begin{equation}
M(u)=k_1 \quad \mbox{for $\epsilon/2 < |u| < 1-\epsilon/2$}
\label{eq:firstorder1}
\end{equation}
and
\begin{equation}
\int_{-\epsilon/2}^{\epsilon/2} M(u) \, du = - k_1 (1-\epsilon).
\label{eq:firstorder2}
\end{equation}
Equations~\ref{eq:firstorder1} and \ref{eq:firstorder2} contain
Equations~\ref{eq:notch1} and \ref{eq:notch2} as a special case ($k_1=0$).

Mask functions $\hat M(x)$ that solve Equations~\ref{eq:firstorder1} and \ref{eq:firstorder2} can always be
decomposed into a sum of a notch-filter function and a $\delta$-function.
For example, a $1 - \delta$-function mask
\begin{eqnarray}
\hat M(x) &=& 1 - \delta(x) \\
\label{eq:delta1}
M(u) &=& \delta(u) - 1 \\
\label{eq:delta2}
{\cal M}(u) &=& H(u) - u
\label{eq:delta3}
\end{eqnarray}
fills the bill.  We can call notch filter masks ``zeroth order'' masks, and
this $1 - \delta$-function mask a ``first order'' mask.

Of course, we can not build this $1 - \delta-$function mask.
The closest we can come is to build a mask that is everywhere equally transparent, but that generates a
phase shift of $\pi$ in the center:
\begin{eqnarray}
\hat M(x) &=& \left\{  \begin{array}{ll}
            1 &\mbox{for $|x| > 1/4$} \\
            -1 &\mbox{for $|x| < 1/4$}. \end{array}
	    \right. \\
\label{eq:phasedisk}
M(u) &=& \delta(u) - \sin(\pi u/2)/(\pi u/2) \\
{\cal M}(u) &\approx& H(u) -  u +   {{\pi^2} \over {72}} u^3  -\dots
\end{eqnarray}
Figure~\ref{fig:calm}d shows ${\cal M}(u)$ for this mask---it approximates the
sawtooth-like exact solution of Figure~\ref{fig:calm}c.
The disk phase mask of \citet{rodd97} is a two-dimensional version of this
approximate solution.

\subsection{The Rest of the Series}

Equation~\ref{eq:demandsonlythat} shows that over some interval, ${\cal M}(u)$, and
therefore $M(u)$ are periodic functions, with period $1$.
We can expand such functions as a Fourier series:
\begin{equation}
P(u)=\sum_{j=0}^{\infty} A_j \sin{2 \pi j x} + B_j \cos{2 \pi j x}
\end{equation}
where $A_j$ and $B_j$ are complex.
This series satisfies Equation~\ref{eq:demandsonlythat}.
However, this solution would completely specify the function over all $x$;
Equation~\ref{eq:demandsonlythat} offers more freedom than that.  We can express a complete
solution as the sum of the above Fourier series and any notch filter function:
\begin{eqnarray}
M(u)&=&N(u) + \sum_{j=1}^{\infty} A_j \cos{2 \pi (j-1) x} + B_j \sin{2 \pi (j-1) x}  \\
\label{eq:complete}
\hat M(x)&=&\hat N(x) + \sum_{j=1}^{\infty}  C_j \, \delta(x - (j-1)) + D_j \, \delta(x+(j-1)),
\end{eqnarray}
where and $C_j=(A_j - i B_j)/2$ and $D_j=(A_j + i B_j)/2$, as long as
\begin{eqnarray}
\int_{-\epsilon/2}^{\epsilon/2} \, M(u) \, du &=& \int_{\epsilon/2}^{1-\epsilon/2} \, M(u) \, du \\
&=& \epsilon A_j \, \mbox{sinc} \, \pi (j-1) \epsilon.
\end{eqnarray}
In other words, all masks that completely remove on-axis light in a coronagraph with
a tophat entrance aperture can be represented as the sum of a notch filter function and
a series of $\delta$-functions located at $x=\dots, -2, -1, 0, 1, 2, \dots$.
So far, we have only considered masks with all $A_j$ and $B_j$ equal to 0,
which we called zeroth order masks, or masks with only $A_1 \ne 0$, which we
called first order masks (there is no $B_1)$.  We can refer to masks with $A_j$ or $B_j \ne 0$, for $j > 1$ as ``higher order'' masks.

We can also consider the situation where the Lyot stop is bigger than the
entrance aperture; $\epsilon < 0$.  In this case, the solution to Equation~\ref{eq:zero}
takes the form
\begin{eqnarray}
M(u)&=&H(u) + \sum_{j=1}^{\infty} A_j \cos{2 \pi j x} + B_j \sin{2 \pi j x}  \\
\hat M(x)&=&\hat H(x) + \sum_{j=1}^{\infty}  C_j \, \delta(x - (j-1)) + D_j \, \delta(x+(j-1)),
\label{eq:biglyot}
\end{eqnarray}
where $\hat H(x)$ is a high-pass filter function, i.e. one where $H(u) = 0 $ for $|u| < 1- \epsilon$.  High-pass filter functions do not have any effect on the image, however;
the functional part of the mask is the series of $\delta$-functions.
One example of a mask that works (approximately) for $\epsilon < 0$ is the 1-D
disk phase mask illustrated above.  It does not suddenly fail if
$\epsilon$ becomes small or negative---though the approximation
to a $\delta$-function degrades towards the edges of the pupil plane.

Masks containing a row of $\delta$-functions may not be buildable, but we could build approximate
versions, like the disk phase mask, and we could build low-pass-filtered versions of some masks 	by convolving
 the mask function with $\mbox{sinc}(2 \pi x)$ for example, (though constructing such masks
 could be tricky!).  

However, all masks in this series except for notch filter masks and first order masks
face the following complication in a broad-band system.  In a real optical system,
$M(x)$ necessarily changes with the wavelength of the light, $\lambda$; for an intensity mask,
$\hat M(x) \approx \hat M(x \lambda/\lambda_0)$, where $\lambda_0$ is the reference wavelength used
for the purpose of defining the diffraction scale.  Other, wavelength-dependent
effects may appear, or be built into the mask, but this dilation is intrinsic to a
classical coronagraph.

Consequently, ${\cal M}(u) \approx {\cal M}(u \lambda_0/\lambda)$.  The notch-filter solution,
for which ${\cal M}(u)$ is constant over a large range in $u$, can trivially
be made to work over a large range in $\lambda$ because of its dilation symmetry.  However, all
other solutions will only work at one wavelength, unless $M(x)$ explicitly changes as a
function of wavelength in some special way other than simple dilation.

\section{OFF-AXIS LIGHT AND A MORE GENERAL CORONAGRAPH}

Now we will assess the impact of a coronagraph on an off-axis source, like an
extrasolar planets located at angle $\theta$ from the optical axis in the plane of the sky.
We can proceed in the context of a more general problem, where $A(u)$ and $L(u)$ can be
any functions at all that we can Fourier transform.

\subsection{The Attenuation Function: Inner Working Angle}

A band-limited mask is what-you-see-is-what-you-get;
a coronagraph with a band-limited mask attenuates the field from a distant source
by a factor of $\hat M(\theta)$ and the Point Spread Function (PSF) in such a coronagraph
does not depend on $\theta$. \citep{kuch02}.
But in general, the PSF in a coronagraph depends on $\theta$,
and clearly, the shapes of many image masks do not directly show how the
mask attenuates off-axis sources.  How can we understand in general what
a coronagraph does to image of an astronomical source?

\citet{kuch02} showed that the high wavenumber components of the mask
function, those with $|u| > 1-\epsilon/2$, do not affect how the mask
interacts with off-axis light.
Consequently, for any mask, we can convolve
$\hat M(x)$ with $\mbox{sinc}(2 \pi (1-\epsilon/2) x)$ and get the same effective mask.
For a notch filter mask, this operation just means looking the
band-limited part of the mask, which as we mentioned, is trivial to interpret.
We will apply a similar idea here to reveal the workings of any image mask.

Consider a point source, providing a field $\delta(x-x_1)$ in the plane
of the sky, and a harmonic mask function $M(u)=m \, \delta(u-u_1)$.
The field after the entrance pupil is
$A(u) \exp(-2 \pi i u x_1)$, and the field in the first image plane is
$\hat A(x-x_1)$.  The field after the image mask is $m \exp(2 \pi i u_1 x) \hat A(x-x_1)$.
The field in the second pupil plane is $m A(u-u_1) \exp(-2 \pi i (u-u_1) x_1)$
The field after the Lyot stop is $m L(u) A(u-u_1) \exp(-2 \pi i (u-u_1) x_1)$.
The final field is $m \exp(2 \pi i u_1 x_1) \mbox{FT}[L(u) A(u-u_1)] * \delta(x-x_1)$.
In other words, the point source response of this particular special coronagraph is an intensity pattern
$|m \, \mbox{FT}[L(u) A(u-u_1)|^2$ centered at $x=x_1$.

We can evaluate the field amplitude
at the center of the pattern simply by replacing the
Fourier transform with $\int_{-\infty}^{\infty} \, du$.  Then we see that the amplitude attenuation provided
by this coronagraph at the center of the image of a point source is
\begin{equation}
\hat F(x_1)  = m \, e^{2 \pi i u_1 x_1} \, [L(u_1) * A(-u_1)] \quad \mbox{for a harmonic mask.}
\end{equation}
The intensity at the center of the Point Spread Function (PSF)
is $|\hat F(x_1)|^2$.  We can call
$\hat F(x_1)$ the coronagraph's amplitude attenuation function and
$|\hat F(x_1)|^2$ the coronagraph's intensity attenuation function.

We can find the attenuation function for any arbitrary mask by expanding the mask
function as $M(u)= \int_{-\infty}^{\infty} \, M(u_1) \, \delta(u - u_1) \, du$.  Then the attenuation function becomes a
Fourier integral:
\begin{eqnarray}
\hat F(x_1) &\equiv& \int_{-\infty}^{\infty} \, M(u) \, e^{2 \pi i u x_1} \, [L(u) * A(-u)]\, du \\
&=& \mbox{FT}(M \cdot (L(u) * A(-u)))(x_1) \\
&=& \left[\hat M * \left (\hat L \cdot \hat A^\dagger \right )\right](x_1),
\label{eq:attenuation}
\end{eqnarray}
where $\dagger$ denotes complex conjugation.  This definition applies for any arbitrary entrance aperture,
image mask, or Lyot stop. The PSF shape may change as a function of $x_1$, the
location of the off-axis source.  However, the center of the PSF is usually the maximum of the PSF,
so the attenuation function measures a critical feature of the mask for the purpose of finding
off-axis point sources, like extrasolar planets.

Equation~\ref{eq:attenuation} shows that while a mask may look to us like it has sharp edges,
in some sense it looks to the sky like it is smooth---convolved with a smoothing kernal
of the form $\hat L \cdot \hat A^\dagger$.
For example, the $\delta-$function mask described above
has an amplitude attenuation function $\hat F(x) = 1 - [\hat L \cdot \hat A^\dagger](x)$.
If the Lyot stop is a tophat function and $\epsilon$ is small, then
$\hat F(x) \approx 1- \mbox{sinc}^2(\pi x/2)$.
If the image mask is band-limited, then $\hat F(x_1)=\hat M(x_1)$.

\begin{figure}
\epsscale{0.9}
\plotone{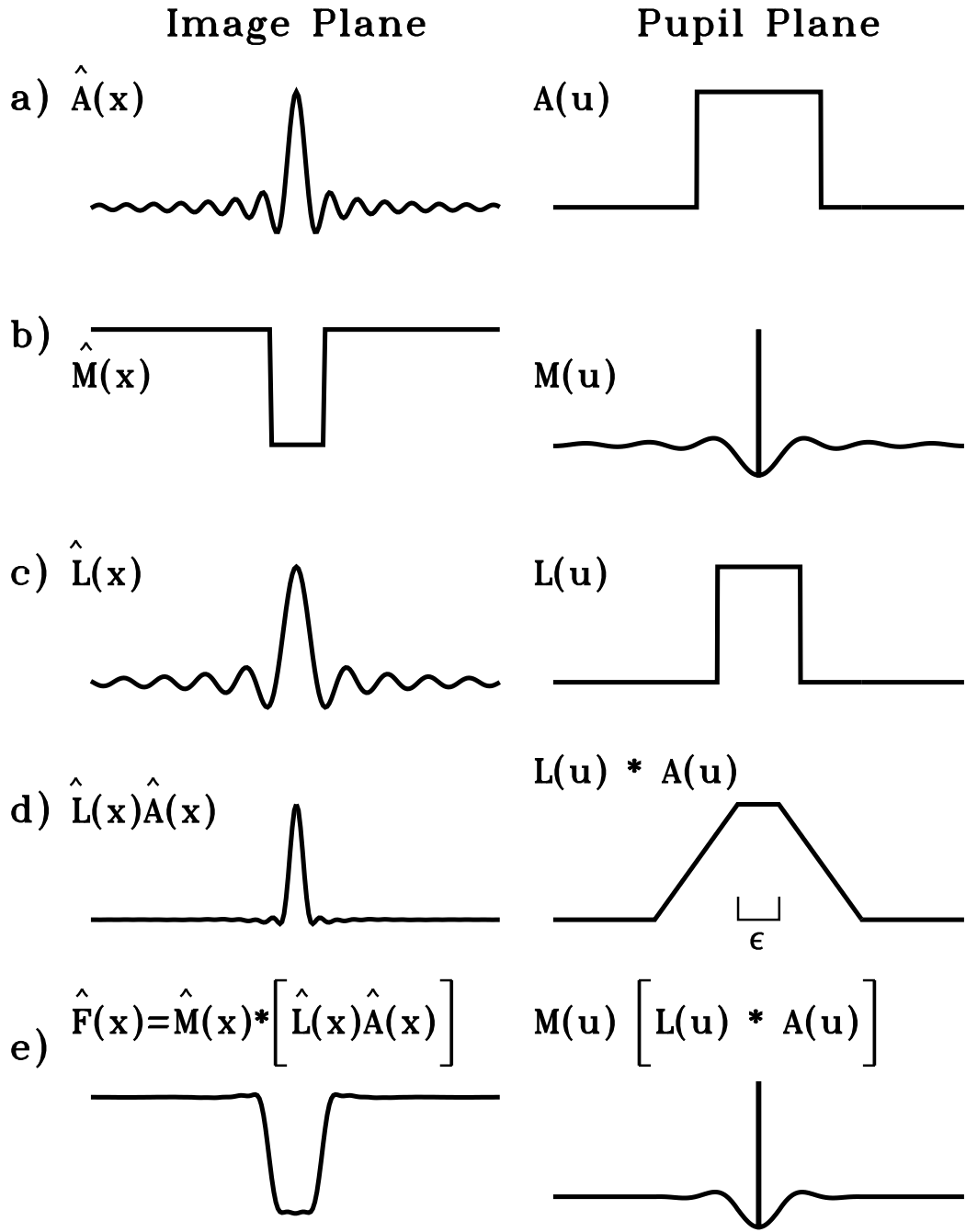}
\caption{Constructing the attenuation function, a smoothed version of the mask that
the planets see. a) entrance pupil and its Fourier conjugate b) image mask and its conjugate
c) Lyot stop and its conjugate d) the smoothing kernal and the filter function e) the amplitude
attenuation function, $\hat F(x)$, and its conjugate.
\label{fig:construct}}
\end{figure}

Figure~\ref{fig:construct} shows how $\hat F(x)$ is constructed for a coronagraph with a tophat
entrance aperture, tophat image mask, and tophat Lyot stop.   The left column shows image plane
quantities, and the right column shows their pupil plane conjugates.
Figure~\ref{fig:construct}a, b, and c show the entrance aperture, $A$, the image mask, $\hat M$, and
the Lyot stop, $L$. Figure~\ref{fig:construct}d shows the convolution of $L(u)$ and $A(u)$ and
the image plane conjugate of this function, $\hat L(x) \hat A(x)$.  These quantities act
respectively as filter function and smoothing kernal. The amplitude attenuation function, $\hat F(x)$,
shown in Figure~\ref{fig:construct}e, is a smoothed version of the mask.

In this example, $L(u)$ * $A(u)$ is a constant for $|u| < \epsilon/2$, and zero for
$|u| > 1 - \epsilon/2$, as shown in Figure~\ref{fig:construct}d.   Consequently, if $M(u)$ were
a band-limited mask function with power only where $|u| < \epsilon/2$, then multiplying $M(u)$ by
$L(u) * A(u)$ would only serve to multiply $M(u)$ by a constant.  In other words, $\hat M(x)$ would
be an eigenfunction of convolution with $\hat L(x) \cdot \hat A(u)$.

\begin{figure}
\epsscale{0.85}
\plotone{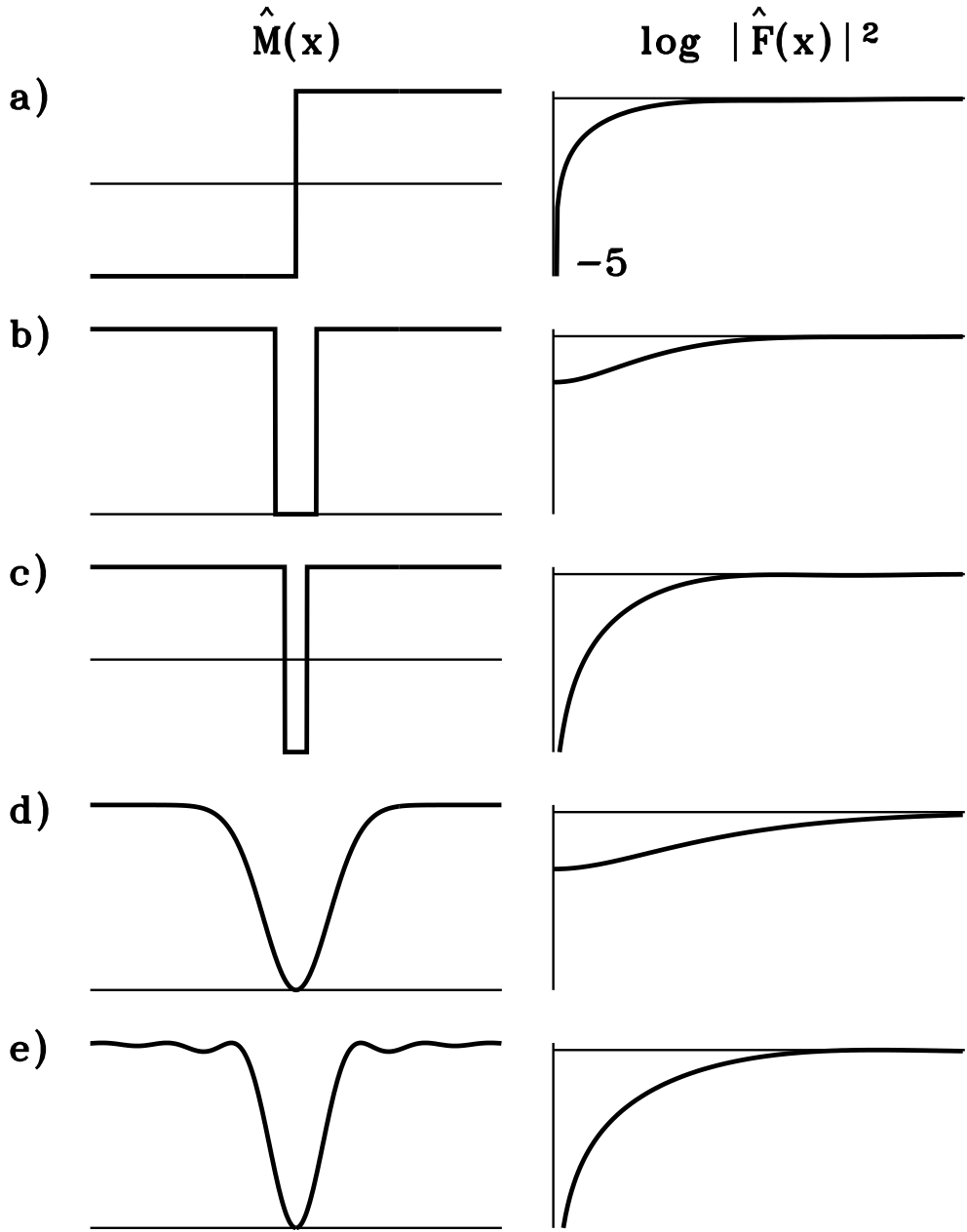}
\caption{Mask functions, $\hat M(x)$, and intensity attenuation functions, $|\hat F(x)|^2$, for
several image masks, calculated assuming the entrance aperture and Lyot stop are identical
tophat functions. a) Phase Knife b) Tophat c) 1-D Phase Disk d) Gaussian e) $1-\mbox{sinc}^2$ (Band-Limited)
\label{fig:compare}}
\end{figure}

Figure~\ref{fig:compare} shows $\hat M(x)$ and $\log |\hat F(x)|^2$ for
several different image masks: a phase knife (Equation~\ref{eq:knife}), a tophat mask, a one-dimensional
disk phase mask (Equation~\ref{eq:phasedisk}), a Gaussian mask, and a band-limited mask
($\hat M(x)=1-\mbox{sinc}^2(\pi \epsilon x/2)$).  Of the masks shown, only
the phase knife and the band-limited mask provide $\hat F(0)=0$.
Of those, only the band-limited mask provides perfect cancellation of on-axis light
and the deep null needed for terrestrial planet finding.


The attenuation function provides a good metric for the inner working angle, $\theta_{IW}$
of a coronagraph.  For example, we could define the inner working angle by
$|\hat F(\theta_{IW})|^2=1/2$.  With this definition, and assuming a tophat Lyot stop with $\epsilon=0$,
$\theta_{IW}=0.58 \lambda/D$ for the $\delta$-function image mask,
and $0.43 \lambda/D$ for a phase knife. For comparison, the
smallest possible inner working angle for a band-limited mask is $\theta_{IW}=0.64 \lambda/D$---except that
for this band-limited mask, $\epsilon=1$, so the coronagraph would have no throughput.

Notice that with a mask that is not a notch filter mask,
apodizing the Lyot stop makes $\hat L$ into a broader function, which increases the
coronagraph's inner working angle.  Likewise, we could push the inner working angle inwards
if we replaced the Lyot stop with a pair of pinholes at $u=\pm 1/2$, for example.
However, in reality, these distinctions probably don't matter, because low order aberrations, like
pointing error, will probably set the inner working angle of a real coronagraph


\subsection{Mask Symmetry and Stellar Leak}

By taking derivatives of Equation~\ref{eq:attenuation}, we can construct a Taylor expansion for $\hat F(x)$
about $x=0$:
\begin{equation}
\hat F(x)  = \sum_{n} {x^n \over {n!}} \int_{-\infty}^{\infty}  M(u) \, (2 \pi i u x_1)^n \, [L(u) * A(-u)] \, du .
\label{eq:taylor}
\end{equation}
Consider a system where $L(u)$ and $A(u)$ have even symmetry.
If $M(u)$ has even symmetry then
$\left. \left({d \over {dx}}\right)^{n} \hat F(x) \right|_{x=0}$ will vanish for all odd $n$.
If $M(u)$ has odd symmetry then
$\left. \left({d \over {dx}}\right)^{n} \hat F(x) \right|_{x=0}$ will vanish for all even $n$.
For example, in a coronagraph where $\hat A(x)$ and $\hat L(x)$ have even symmetry, if
we use any image mask with odd symmetry, the on-axis
final image field from an on-axis point source will be zero.

Mask functions, $\hat M(x)$, with odd symmetry necessarily
become negative, so they require manipulating the phase of the beam.
The simplest such mask is the phase-knife
coronagraph \citep{abe01}, for which
\begin{equation}
\hat M(x) = \left\{  \begin{array}{ll}
            1 &\mbox{for $x > 0$} \\
            0 &\mbox{for $x=0$} \\
            -1 &\mbox{for $x < 0$}. \end{array}
	    \right.
\label{eq:knife}
\end{equation}
A phase knife consists of a half-plane of glass joined to a half-plane of
phase-retarding material so that the seam falls on the optical axis.  We may be able to
make a more practical version of this mask by laying
an opaque strip over the seam so that the mask function becomes
\begin{equation}
\hat M(x) = \left\{  \begin{array}{ll}
            1 &\mbox{for $x > s$} \\
            0 &\mbox{for $|x| < s$} \\
            -1 &\mbox{for $x < -s$}. \end{array}
	    \right.
\end{equation}
where $s$ is one or two diffraction widths.  This variation retains the mask's odd symmetry,
but removes the seams from direct illumination.



Though $\hat F(x=0)=0$ for the phase knife, the mask does not eliminate all on-axis light.
The on-axis source may produce an image with zero central intensity, but with
bright wings that can overwhelm the image of an off-axis planet.
A perpendicular pair of phase knives, like the four-quadrant phase mask \citep{roua00, riau01, riau03, lloy03}
does a better job than a single phase knife at reducing these wings at the cost of some search area
and also exactly satisfies Equation~\ref{eq:zero} for a circular entrance aperture and circular Lyot stop.

\begin{table}[h]
     \caption{Null Depth}
      \medskip
\begin{tabular}{lcc} \hline\hline
Mask                           & Intensity Leak   & Pointing Requirement \\
                               & On Axis          & (fraction of $\theta_{IW}$) \\ \hline\hline
Tophat                         & $\theta^0$       & -      \\
Disk Phase Mask                & $\theta^0$       & -      \\
Phase Knife                    & $\theta^2$       & 0.0001 \\
Four-Quadrant Phase Mask       & $\theta^2$       & 0.0001 \\
All Masks With Odd Symmetry    & $\theta^2$       & 0.0001    \\
Notch Filter                   & $\theta^4$       & 0.01   \\
Band-limited                   & $\theta^4$       & 0.01   \\
Gaussian                       & $\theta^4$       & 0.01${}^{a}$   \\
Achromatic Dual Zone           & $\theta^4$       & 0.01${}^{a}$   \\
\hline\hline
\end{tabular}
\tablenotetext{a}{Assuming appropriate pupil stops that the zeroth order leak is negligible.}
\label{tab:table1}
\end{table}

However, the wings of the image of an on-axis point source are not the only drawback to
the phase-knife mask.  Real stars do not provide perfect on-axis point sources; stars have finite
angular size and telescopes can not point at them perfectly accurately \citep[e.g.][]{riau01,kuch02,kuch03,lloy03}.
We must consider the  intensity leak from slightly off-axis starlight, which we can
approximate by $|\hat F(\theta)|^2$, where $\theta$ is the angle off axis.

A common misconception is that disk phase masks are especially sensitive to pointing error.
However, any coronagraph has a pointing error tolerance related to its inner working angle;
the smaller the inner working angle, the tighter the tolerance.  Disk plase masks have
small inner working angles, so they have tight pointing tolerances.  But at a given inner working
angle, disk phase masks perform well.

In general, the leak increases as some power of $\theta$.  Equation~\ref{eq:taylor} shows that
the intensity leak from a mask with odd symmetry, like a phase-knife mask, is ${\cal O}(\theta^2)$.
A notch filter mask has even symmetry, so the intensity leak from a notch-filter mask is ${\cal O}(\theta^4)$.
The circular phase mask \citep{rodd97} and the Gaussian mask have
$F(x=0) \neq 0$, so they leak at ${\cal O}(\theta^0)$, though the zeroth order leak
can be made negligible.
All masks of containing $\sin$ terms in Equation~\ref{eq:complete} produce an
${\cal O}(\theta^2)$ intensity leak.

Viewed at quadrature, reflected visible light from the Earth is $2 \times 10^{-10}$ times as bright as
direct light the Sun.  We do not need to suppress the starlight to quite this level in the center of the
image plane since the planet appears typically a few diffraction widths from the star in the wings of
the stellar leakage, which we can control with an apodized Lyot stop if necessary.
A reasonable assumption might be that for terrestrial planet finding, we can tolerate up to
$\hat F(\Delta \theta)=10^{-8}$, where $\Delta \theta$ is the effective pointing error (which may contain
some contribution from the star's finite angular size).

Given this assumption, we must center the star on the mask to an accuracy of roughly
$\Delta \theta=10^{-8/\beta} \theta_{IW}$, where the
mask's leak is ${\cal O}(\theta^{\beta})$, and $\theta_{IW}$ is the coronagraph's inner working angle.
For example, a fourth-order intensity leak, or a fourth-order ``null'', to use the
language of interferometry, translates into a pointing requirement
of $\Delta \theta \lesssim \theta_{IW}/100$.  Table~1 summarizes the leading order
intensity leaks from the masks discussed in this paper.

\subsection{Other Directions: Apodized Entrance Apertures}

We found a complete solution to the problem of removing on-axis light in a one-dimensional coronagraph
with an un-apodized entrance aperture.  However, several coronagraph designs, classical and otherwise, use
apodized entrance pupils \citep{kasd03} or specially shaped pupils \citep{sper01, kasd03, vand03a, vand03b}.
Other designs use a pair of shaped mirrors to generate an apodized beam \citep{guyo03, trau03}.
These designs aim to suppress the wings of the PSF so far that the edges of the image mask
are not illuminated by the image of an on-axis source, so a simple tophat mask suffices to
remove the on-axis light to the necessary level.  Achieving this suppression generally requires
numerical optimization to find an approximate solution to Equation~\ref{eq:zero}.
Coronagraphs with apodized entrance apertures, like coronagraphs with apodized Lyot stops, are
especially robust to low-order aberrations.


We can also find new exact
solutions to Equation~\ref{eq:zero} using an apodized entrance aperture and a phase mask
\citep{aime02, saf03}.  However, these exact solutions
require $\hat M(x)$ to vary in a complicated way with wavelength.  One design
that provides a self-adjusting $\hat M(x, \lambda)$ to be combined with an apodized entrance aperture
is the achromatic dual-zone phase coronagraph \citep{soum03}.  Achromatic dual-zone phase masks
can be designed with ${\cal O}(\theta^4)$ leak.  Combining image masks and apodized entrance pupils remains
relatively unexplored territory---though perhaps such designs shold not be called classical coronagraphs.

\section{CONCLUSION}

According to Fraunhofer diffraction theory, uncountably many non-trivial image mask
designs can completely remove the light from an on-axis source in a classical coronagraph with an
un-apodized entrance aperture.  We organize these designs using a trigonometric series solution to the
equation $L \cdot (M*A)=0$.  Notch filter masks are the zeroth order solutions, and the only solutions
that are trivially achromatic in our approximation over a large bandwidth.
Higher order solutions can be decomposed into the sum of a notch filter function and a
series of $\delta(x)$ functions; the
the disk phase mask \citep{rodd97} approximates one term in the sum.

We defined the mask amplitude attenuation function, $\hat F(x)$, as the amplitude at center of the final image
of a point source located at angle $x$ off-axis in the plane of the sky.  This function, which we showed is a
smoothed version of the mask function, provides an easy way to judge the null depth and inner
working angle for non-band-limited masks.  Through this analysis, we showed that
only masks with even symmetry can provide the null depth necessary for terrestrial planet searches;
for example, masks based on a phase knife do not provide the necessary null depth.

We focused on a one-dimensional coronagraph with an un-apodized entrance aperture, though other designs
may prove useful. For example, the full two-dimensional problem may yield surprises.
Some coronagraph designs use apodized entrance pupils---though classical coronagraphs with apodized
entrance pupils generally require $\hat M(x)$ to vary with wavelength in a complicated way.
It may also be possible to gain benefit by cascading multiple image plane masks.
But neglecting these avenues, our analysis shows that our existing library of one-dimensional
image mask designs is complete.

\acknowledgments

Thanks to David Spergel,  Wes Traub, and Ann Bragg for comments on this manuscript, and the folks at the Leiden Coronagraphy Workshop for inspiration.
M.J.K. acknowledges the support of the Hubble Fellowship Program of the Space Telescope Science Institute.

\end{document}